\documentclass[runningheads,a4paper]{llncs}
\usepackage{graphicx}

\begin{document}

\title{Prolog Visualization System Using Logichart Diagrams}
%
%
\author{Yoshihiro Adachi}
%
%
%
\institute{Toyo University, 2100 Kujirai Kawagoe Saitama 3508585, Japan,\\
\email{adachi@toyonet.toyo.ac.jp}}

\maketitle              
\setcounter{page}{8}

\begin{abstract}
We have developed a Prolog visualization system that is intended 
to support Prolog programming education. 
The system uses Logichart diagrams to visualize Prolog programs. 
The Logichart diagram is designed to visualize the Prolog execution 
flow intelligibly and to enable users to easily correlate the Prolog 
clauses with its parts. 
The system has the following functions. 
(1) It visually traces Prolog execution (goal calling, success, and failure) 
on the Logichart diagram. 
(2) Dynamic change in a Prolog program by calling extra-logical 
predicates, such as `assertz' and `retract', is visualized 
in real time. 
(3) Variable substitution processes are displayed in a text widget
in real time. 
\end{abstract}

\section{Introduction}
Prolog is a representative programming language for 
introductory education in AI programming. 
It has several characteristic mechanisms, including powerful 
pattern matching (unification, in Prolog terminology), 
automatic backtracking, and meta-programming. 
However, the implementation of these mechanisms is unique, 
so it is difficult for beginners to learn Prolog, especially if 
they have experience in procedural programming languages like 
C and BASIC. 
Extra-logical predicates, such as `assertz' and `retract', 
enable knowledge data to be altered dynamically and 
meta-programs to be created but they can also make Prolog programs 
difficult to understand and debug. 

Visualization using program diagrams can effectively 
facilitate the understanding and debugging of programs.
The Transparent Prolog Machine, a well-known Prolog 
visualization system\cite{TPM1,TPM2}, displays the structure of 
a pure Prolog program as a tree with AND/OR branches 
(an AND/OR tree) and depicts the states of the various goals 
as symbols at its nodes. 
Other visualization and debugging systems for Prolog and 
(constraint) logic programming languages 
(e.g. \cite{Tamir,Carro,Cameron}) also use AND/OR trees. 
However, it is not easy to correlate the content of a 
Prolog program with that of its corresponding AND/OR tree 
because the structure of the clauses of the Prolog program 
and their representations in the AND/OR tree are different. 

\begin{figure*}[hbt]
\begin{center}
\includegraphics[width=120mm]{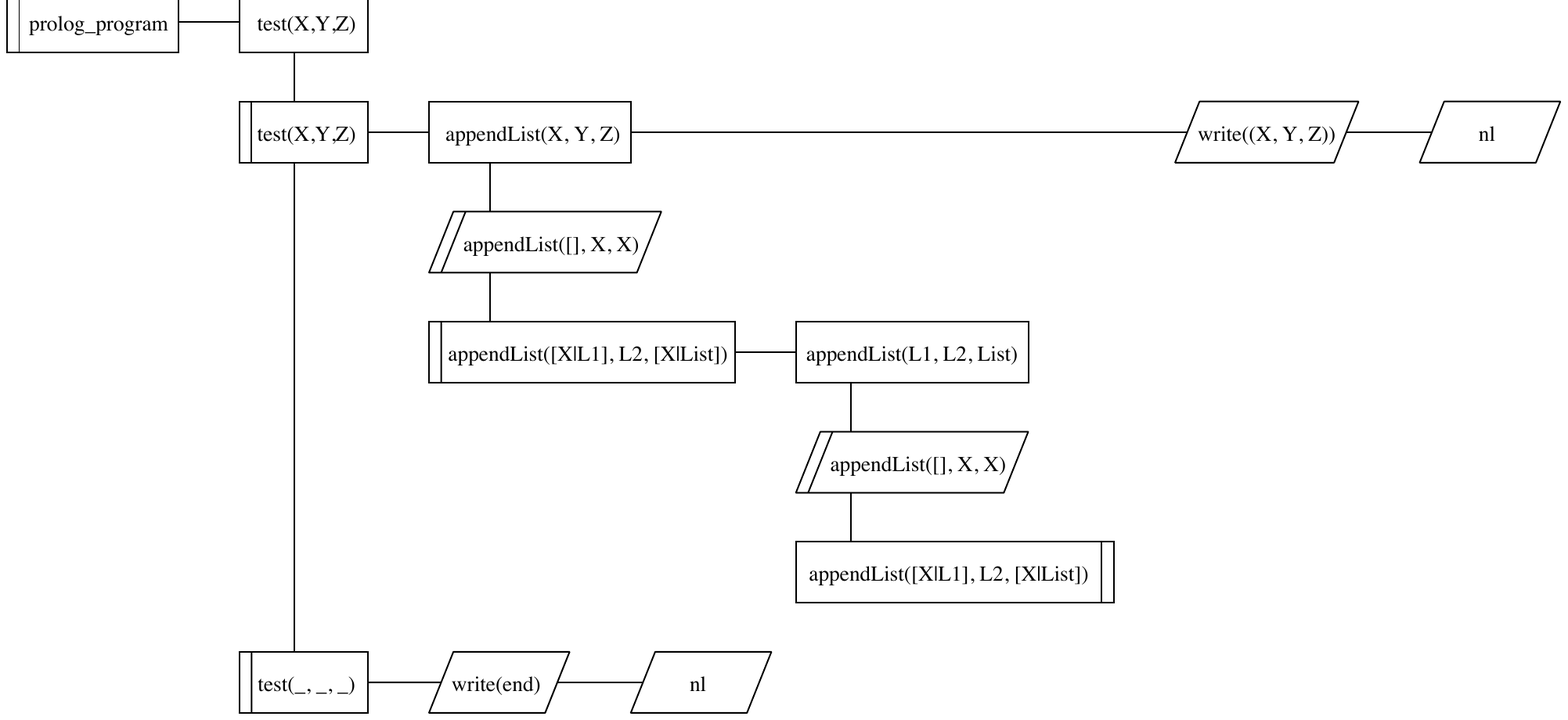}
\caption{Logichart diagram}
\label{logichart}
\end{center}
\end{figure*}

This paper is the first report that describes a Prolog 
visualization system, which we have implemented in SICStus Prolog 
\cite{Sicstus} to support Prolog programming education. 
The system uses Logichart diagrams to visualize Prolog programs. 
Logichart is a program diagram description language that 
we developed to help visualize the execution flow of 
Prolog programs \cite{IFIP98,ENTCS,LED2007}. 
A Logichart diagram has a tree-like structure, as shown 
in Fig. \ref{logichart}, with the following two features. 
(1) The head and body goals that compose each clause are 
aligned horizontally, and (2) a calling goal and 
the heads of the clauses that it calls are aligned 
vertically. 
Feature (1) gives clauses in a Prolog program and their 
representations in the corresponding Logichart diagram 
a similar structure so that it is easier to see the 
correspondences between them. 
Feature (2) makes it easier to understand the relationships between 
related clauses, because they are vertically adjacent. 

The system has three functions: 
(1) it animates Prolog execution (goal calling, success, and failure) 
on the Logichart diagram, 
(2) it visualizes a dynamic change in a Prolog program in real time 
by calling extra-logical predicates, such as `assertz' and `retract', 
and (3) it displays variable substitution processes in a text widget 
in real time. 

\section{Logichart diagrams and Logichart-AGG}
Logichart diagrams have been developed to represent 
computation, which is the response of a Prolog program to 
a query, as an intelligible diagram \cite{IFIP98,ENTCS,LED2007}. 
A Prolog program is visualized by using Logichart diagrams 
as follows. 
For goal sequences `G$_1$, G$_2,\cdots,$ G$_n$' of a user's query, 
the system adds the clause `prolog\_program :- G$_1$, G$_2$, $\cdots,$ G$_n$.' 
to the program. 
The node labeled `prolog\_program' corresponding to the head of this clause 
is the root node of the Logichart diagram. 
The head and body goals composing a clause are horizontally aligned 
from left to right according to the Prolog syntax.
However, a goal and clauses, which have heads that can be unified with it, 
are vertically aligned from top to bottom in the same order 
as the clauses in the Prolog program. 
In this manner, the Logichart is defined based on the Prolog syntax 
and the evaluation rule of Prolog interpreters (leftmost derivation, 
depth-first search). 
As a result, a Logichart diagram is relatively easy to understand, 
and correspondence with the source Prolog program is 
clearly presented. 
It must be noted that the Logichart diagram includes an execution tree 
(whose root node is labeled `prolog\_program') for the user's query 
as a subtree. 

Figure~\ref{logichart} shows a Logichart diagram 
that corresponds to the Prolog program shown below 
and the query `?- test(X,Y,Z).'.

\begin{verbatim}
   test(X,Y,Z) :- appendList(X,Y,Z),
        write((X,Y,Z)),nl.
   test(_,_,_) :- write(end),nl.
   appendList([],X,X).
   appendList([X|L1],L2,[X|List]) :- 
        appendList(L1,L2,List).
\end{verbatim}

\begin{figure*}[hbt]
\begin{center}
\includegraphics[width=40mm]{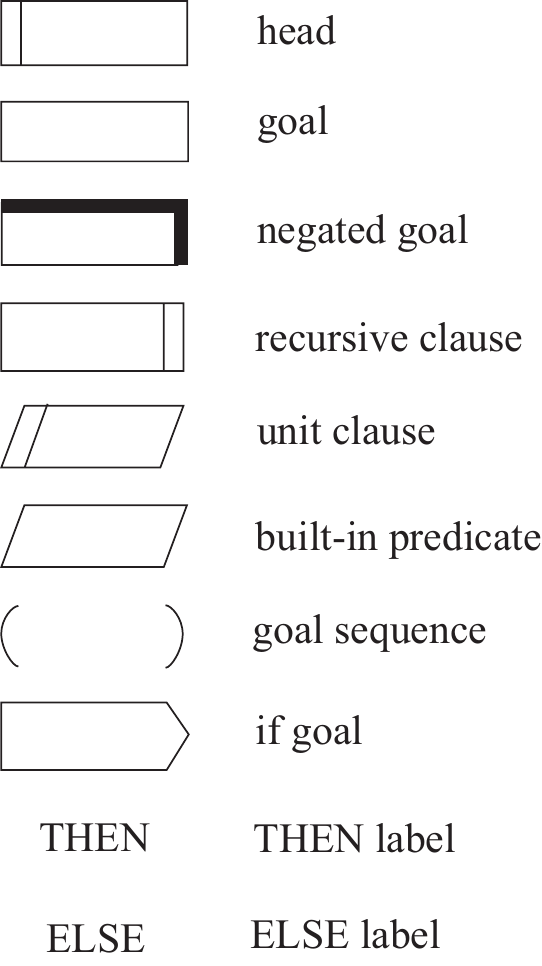}
\caption{Node labels of Logichart diagrams}
\label{labels}
\end{center}
\end{figure*}

Node labels used in Logichart diagrams are shown in Fig. \ref{labels}. 
The heads, body goals, built-in predicates, and user-defined predicates 
are clearly and distinctively depicted using these node labels. 
The label `recursive clause' enables depicting a Prolog program 
including recursive goals within a finite-area Logichart diagram. 

\begin{figure}[htb]
\begin{center}
\includegraphics[width=90mm]{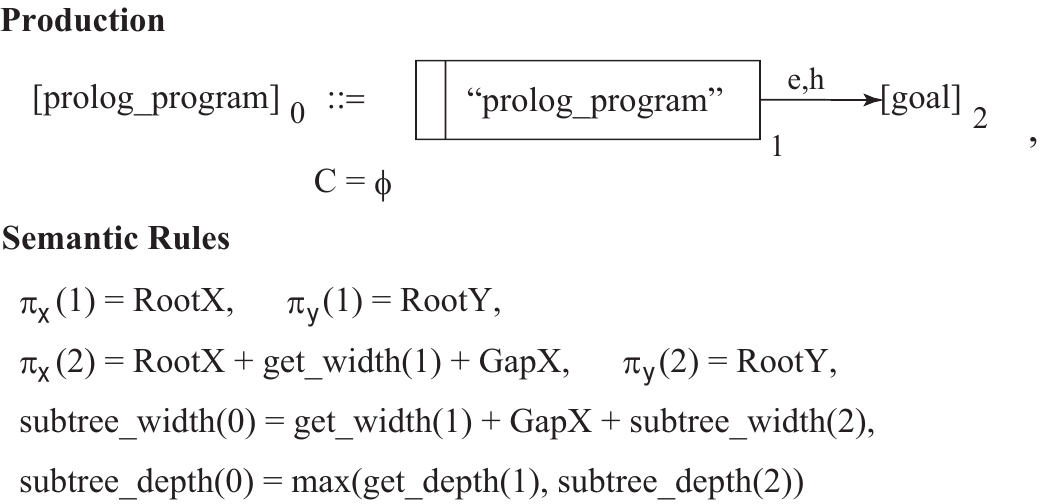}
\caption{Rules used to rewrite initial node `[ prolog\_program ]'}
\label{prod1}
\end{center}
\end{figure}

\begin{figure}[htb]
\begin{center}
\includegraphics[width=90mm]{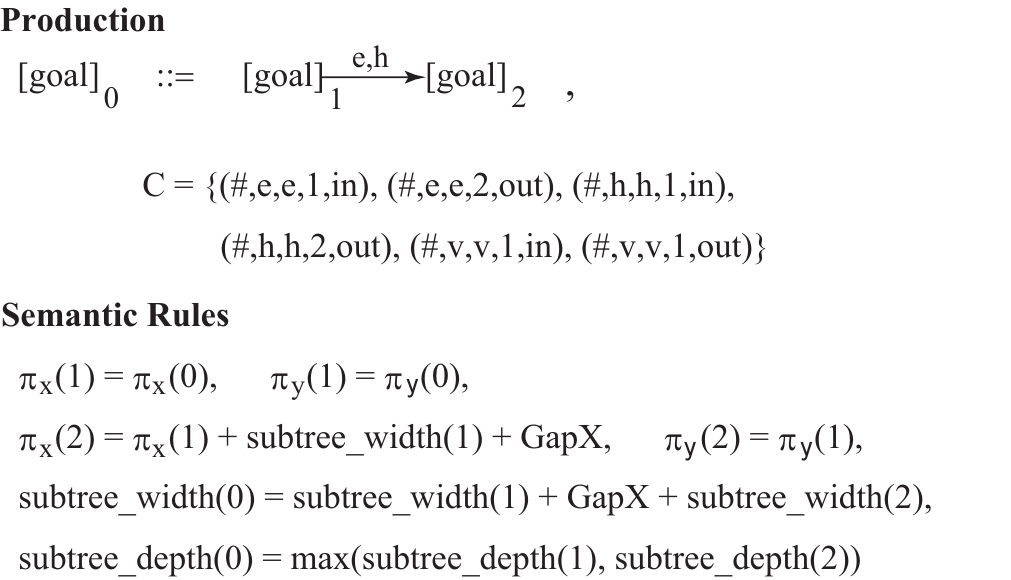}
\caption{Rules formalized for `and' operation on Prolog goals}
\label{prod2}
\end{center}
\end{figure}

We formalized Logichart-AGG \cite{IFIP98,ENTCS,LED2007}, 
an attribute graph grammar defined for specifying the syntax and 
layout rules of Logichart diagrams. 
It consists of a context-free graph grammar whose productions 
are formalized to specify the graph-syntax rules of Logichart diagrams. 
It also uses semantic rules, which are defined so that 
they extract the layout information needed to display 
a Logichart diagram as the attributes attached to node labels. 
The semantic rules are formalized so as to obtain Logichart diagrams 
for a minimum-area layout under a specific layout constraint. 
The Logichart-AGG specifications are very concise and consist of 
13 productions associated with 88 semantic rules.
Some of the productions and their associated semantic rules 
in the Logichart-AGG are illustrated in Figs. \ref{prod1} 
to \ref{prod4}. 

Figure \ref{prod1} shows the production and semantic rules to rewrite 
the initial node `[ prolog\_ program ]'.
These rules are formalized to represent queries given in the Prolog syntax, 
and the nonterminal node `goal' in the right-hand-side graph 
corresponds to the query. 
A graph that is isomorphic to the right-hand side of Production 1 is 
derived by applying this production to the initial node. 
Semantic rules $\pi_x(1)={\rm RootX}$ and $\pi_y(1)={\rm RootY}$ mean
that the x-coordinate of node `1' is `RootX' and that the y-coordinate of 
node `1' is `RootY'. 
Semantic rule $\pi_x(2)={\rm RootX}+{\rm get\_width(``prolog\_program")}
+{\rm GapX}$ means that the x-coordinate of node `2' is equal to `RootX' 
plus the width of the node labeled ``prolog\_program'' plus the horizontal 
gap `GapX'. 
Semantic rule $\pi_y(2)= {\rm RootY}$ means that the y-coordinate of 
node `2' is `RootY'. 
The root node ``prolog\_program'' and the subdiagram derived from the 
nonterminal node `2' labeled `goal' are aligned with a horizontal 
separation of `GapX' by these semantic rules. 

The production and semantic rules shown in Fig. \ref{prod2} are 
as formalized for the `and' operation on Prolog goals. 
A node labeled `goal' is replaced with a graph that is isomorphic 
to the right-hand side of Production 2 by applying this production. 
Semantic rule $\pi_x(2)=\pi_x(1)+{\rm subtree\_width(1)}+{\rm GapX}$ 
means that the x-coordinate of node `2' is equal to the x-coordinate 
of node `1' plus the width of the subdiagram derived from node `1' plus 
the horizontal gap `GapX'. 
Therefore, goals connected by the operator `and' are aligned with 
a separation of `GapX' horizontally. 

\begin{figure}[htb]
\begin{center}
\includegraphics[width=87mm]{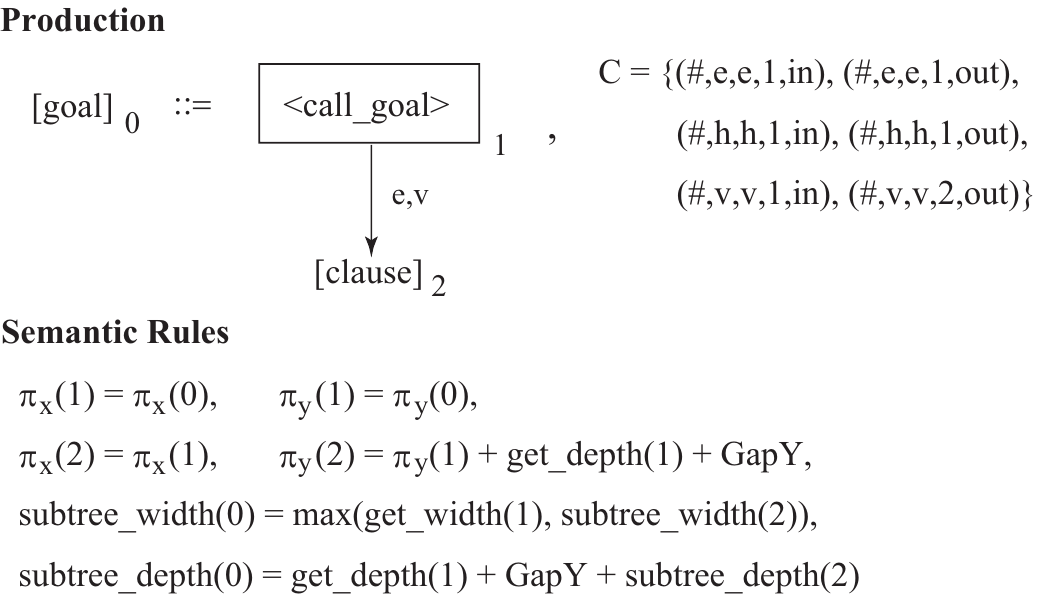}
\caption{Rules formalized for call of goal}
\label{prod4}
\end{center}
\end{figure}

The production and semantic rules shown in Fig. \ref{prod4} are formalized 
for the call of a goal. 
The semantic rule $\pi_y(2)=\pi_y(1)+{\rm depth(1)}+{\rm GapY}$ means 
that the y-coordinate of node `2' is equal to the y-coordinate of 
node `1' plus the depth of node `1' plus 
the vertical gap `GapY'. 
Therefore, a calling goal and the clause heads of the goals called by it 
are aligned with a separation of `GapY' vertically. 

Implementing the Prolog visualization system in complete accord 
with Logichart-AGG guarantees that, for any correct Prolog 
program, the corresponding Logichart diagram is displayed (completeness) 
and any Logichart diagram displayed by the system is valid 
for Logichart-AGG (soundness) \cite{VLC2005}.

\section{Prolog visualization system}
The Prolog visualization system is implemented in SICStus Prolog with 
the Tkl/Tk library \cite{Sicstus} and is intended to support Prolog 
programming education. 

A user edits and then saves a Prolog program by using the system's 
text editor or other editors. 
He/she then inputs the Prolog program into the system. 
Next, he/she inputs a query into the system via a query-inputting 
window. 
A Logichart diagram for the Prolog program and the query 
is then depicted in a canvas widget. 
Figure \ref{text} shows the system's text editor, and Fig. \ref{query} shows 
the query-inputting window. 

\begin{figure}[htb]
\begin{minipage}{.5\linewidth}
\begin{center}
\resizebox{57mm}{!}{\includegraphics{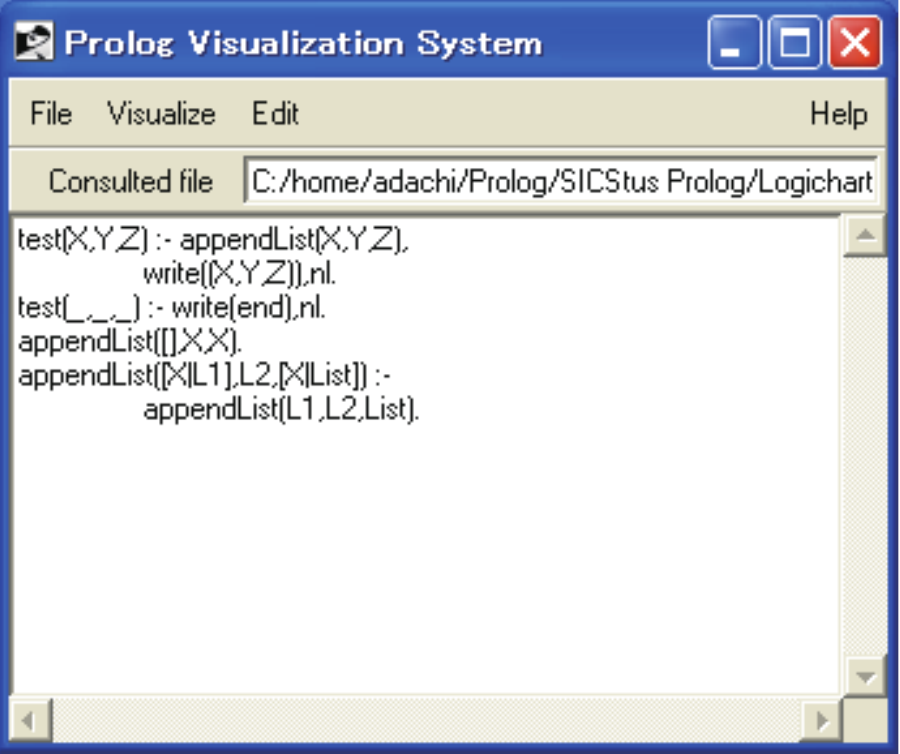}}
\end{center}
\caption{System's text editor}
\label{text}
\end{minipage}
\setcounter{figure}{6}
\begin{minipage}{.5\linewidth}
\begin{center}
\resizebox{33mm}{!}{\includegraphics{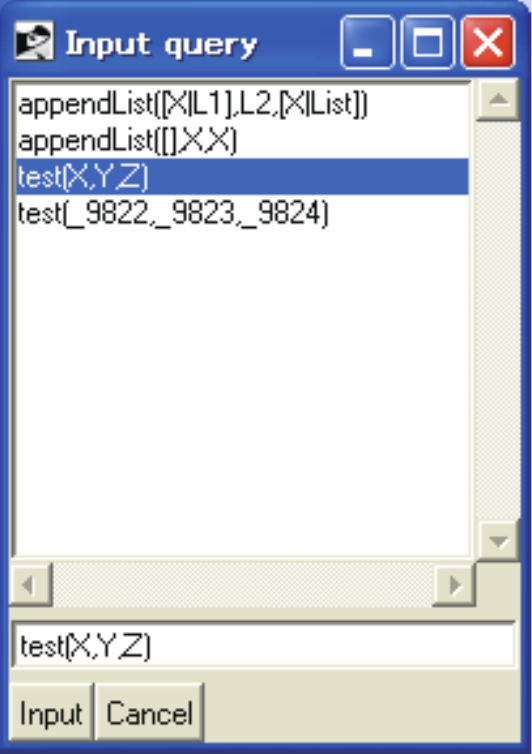}}
\end{center}
\caption{Query-inputting window}
\label{query}
\end{minipage}
\end{figure}

The system has some features specifically designed to support 
Prolog programming learning. 

\subsection{Visual trace of Prolog execution}
The system visually traces Prolog execution on a Logichart diagram 
in real time. 
A tracer implemented using a meta-interpreter technique displays 
the goal-execution process on the Logichart diagram; 
the color of each node changes depending on the goal state, 
i.e., the node corresponding to a goal that has been called 
becomes green, a goal that has succeeded becomes blue, and 
a goal that has failed becomes red. 

\begin{figure*}[hbt]
\begin{center}
\includegraphics[width=100mm]{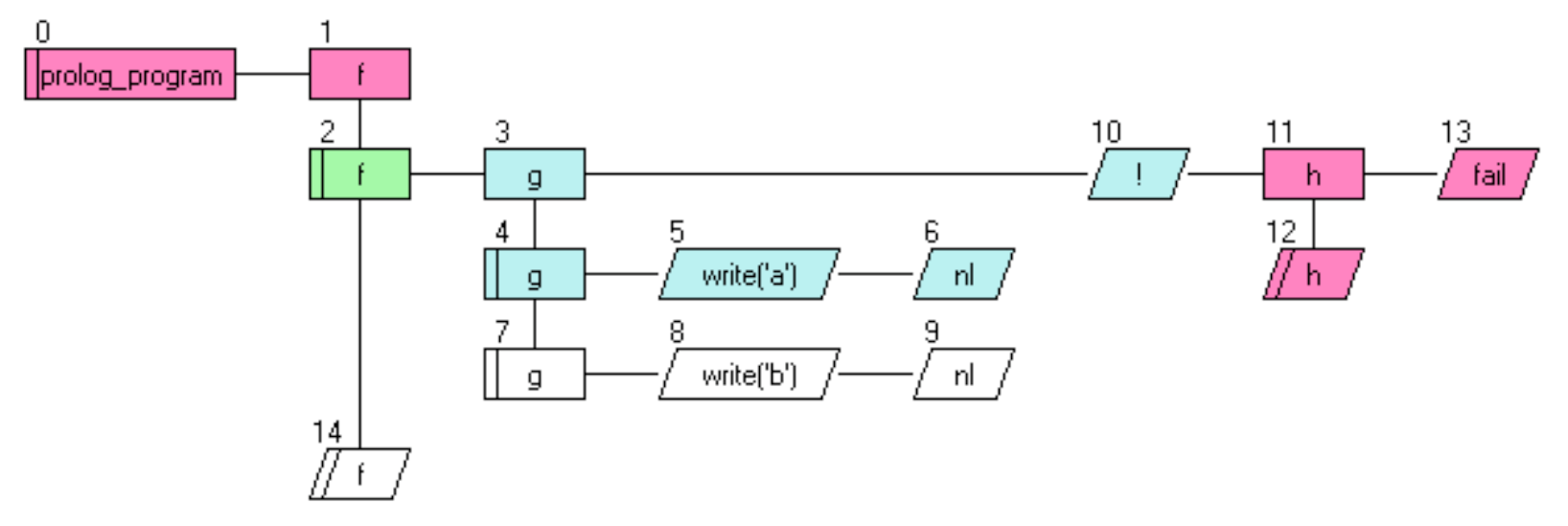}
\caption{Visual trace of Prolog execution}
\label{cut}
\end{center}
\end{figure*}

The tracer has two modes, i.e., one-step and automatic. 
In the one-step mode, each step of goal calling is executed and 
visualized in the Logichart diagram whenever the `go' button is 
pressed, while in the automatic mode the query is executed completely 
and the final state is visualized in the Logichart diagram. 
The effect of cut (`!') on a program execution flow is clearly and 
intelligibly visualized in the one-step mode on a Logichart diagram. 
Figure \ref{cut} shows a screen shot of a visual trace of 
the Prolog program shown below and the query `?- f.'.

\begin{verbatim}
   f :- g, !, h, fail.
   f. 
   g :- write(a),nl.
   g :- write(b),nl.
   h.
\end{verbatim}

If the goals specified in a query are satisfied, and if 
variables are included in the query, then the system displays 
a messageBox window that asks whether the system executes 
backtracking or not. 
If the user clicks the `yes' button, the system executes 
backtracking and the backtracking process is visualized 
on the Logichart diagram.

\subsection{Dynamic change in the Logichart diagram}
\begin{figure*}[hbt]
\begin{center}
\includegraphics[width=120mm]{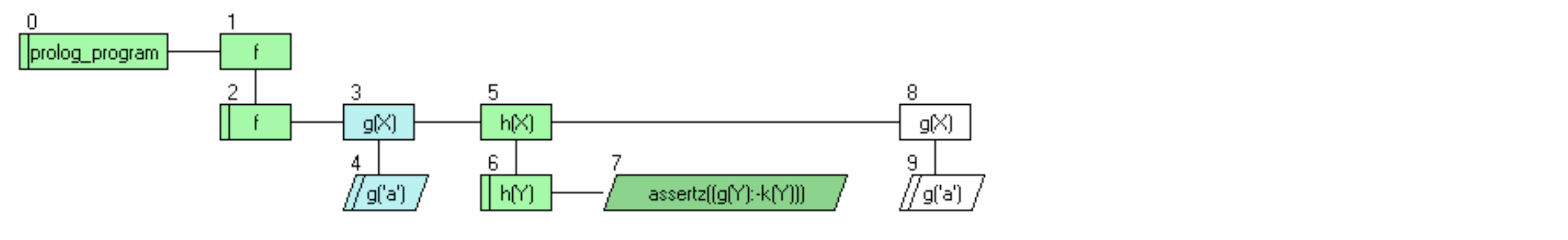}
\caption{Logichart diagram before change}
\label{before}
\end{center}

\begin{center}
\includegraphics[width=120mm]{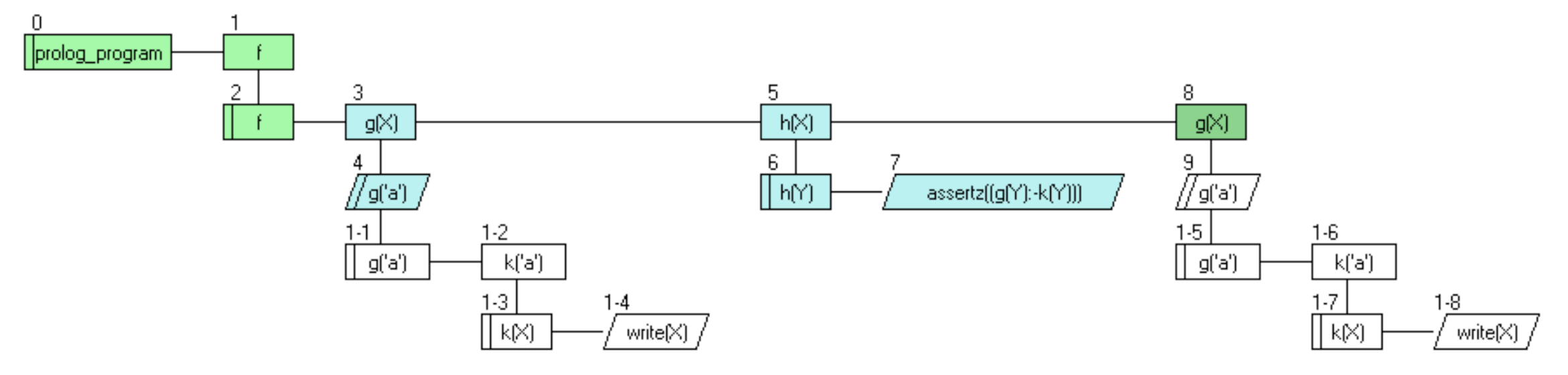}
\caption{Logichart diagram after change}
\label{after}
\end{center}
\end{figure*}

Dynamic changes in a Prolog program by calling extra-logical 
predicates, such as `asserta', `assertz', and `retract', are 
visualized in real time using a Logichart diagram. 
Figures \ref{before} and \ref{after} show a dynamic change 
in the Prolog program shown below and the query `?- f.' 
by calling `assertz((g(Y) :- k(Y)))'. 

\begin{verbatim}
   f :- g(X), h(X), g(X).
   g(a).
   h(Y) :- assertz((g(Y) :- k(Y))).
   k(X) :- write(X).
\end{verbatim}

\begin{figure*}[hbt]
\begin{center}
\includegraphics[width=120mm]{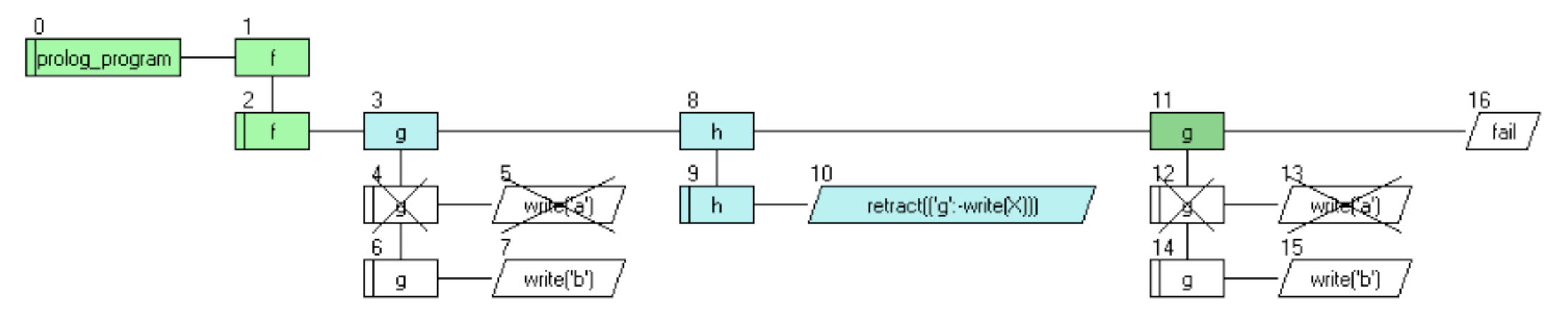}
\caption{Logichart diagram after calling `retract((g :- write(X)))'}
\label{retract}
\end{center}
\end{figure*}

Clauses retracted by calling `retract' are not eliminated but 
depicted with crosses in a Logichart diagram. 
For the Prolog program shown below and the query `?- f.', 
a screen shot of a Logichart diagram after calling 
`retract((g :- write(a)))' is shown in Fig. \ref{retract}. 

\begin{verbatim}
   f :- g, h, g, fail.
   g :- write(a).
   g :- write(b).
   h :- retract((g :- write(X))).
\end{verbatim}

If the nodes corresponding to those of the retracted clauses 
are eliminated from the Logichart diagram, the backtracking process 
in the Logichart diagram becomes impossible to visualize. 
Depicting retracted clauses with crosses enables visualizing 
the backtacking process in the Logichart diagram. 

\subsection{Display of variable substitutions}
\begin{figure*}[hbt]
\begin{center}
\includegraphics[width=50mm]{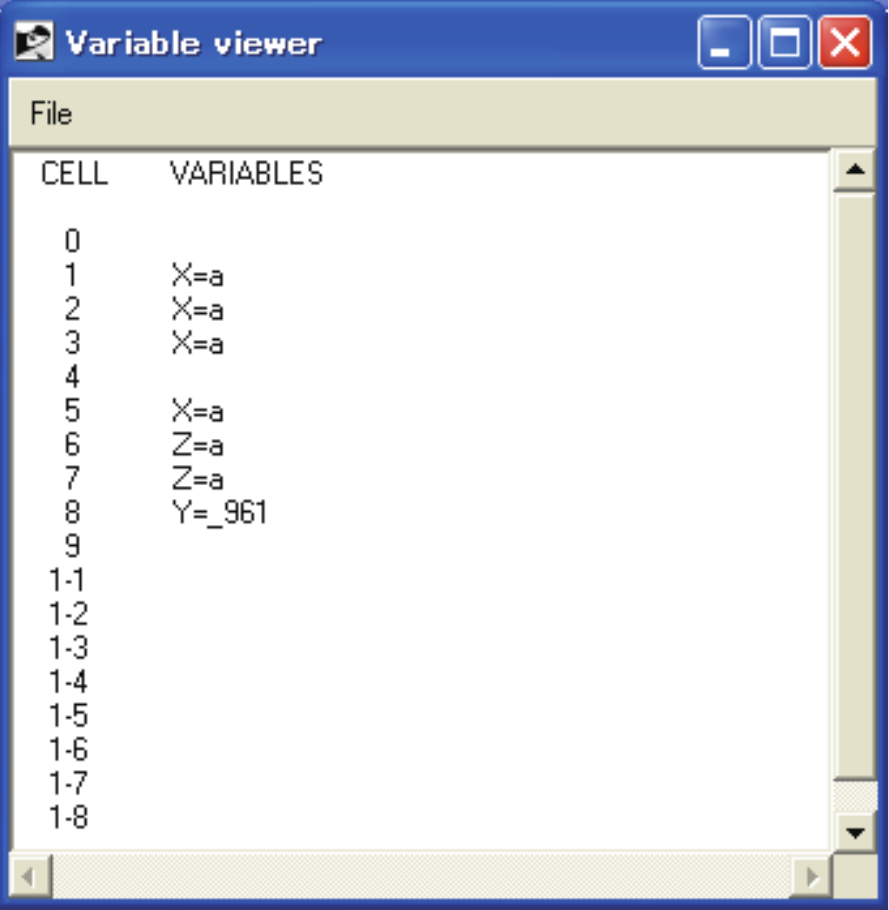}
\caption{Text widget displaying node identifiers and variable substitution}
\label{variable}
\end{center}
\end{figure*}

The variable-substitution process is displayed in a text widget
in real time. 
Figure \ref{variable} shows a text widget displaying variable 
substitution for each node. 
We are now implementing a function to display variable-substitution 
information within the Logichart diagram.

\section{Conclusions}
We presented a Prolog visualization system that is implemented 
in complete accord with Logichart-AGG. 
Logichart diagrams make it easy to understand the Prolog execution flow.
A remarkable feature of the system is that it visualizes 
the dynamic alteration of a Logichart diagram by calling 
extra-logical predicates, such as `assertz' and `retract'. 

We are currently developing fine- and coarse-grained 
(Logichart) diagrams as proposed in \cite{TPM1} to visualize 
and navigate in large execution trees. 
These will help to develop the Prolog visualization system 
into a practical Prolog developing environment. 
The usefulness of our Prolog visualization system for Prolog 
programming education needs to be empirically investigated.

\end{document}